\RequirePackage{lineno}
\documentclass[aps,prl,twocolumn,superscriptaddress]{revtex4}

\usepackage{graphicx}
\usepackage{bm}
\usepackage{amsmath}
\usepackage{amssymb}
\usepackage{hyperref}

\begin{document}
\title{Dramatic effect of a transverse electric field on frictional properties of graphene}

\author{Zhao Wang}
\email{zw@gxu.edu.cn; zhao.wang@tuwien.ac.at}

\affiliation{Department of Physics, Guangxi University, Nanning 530004, China}
\affiliation{Institute of Materials Chemistry, TU Wien, A-1060 Vienna, Austria}
\altaffiliation{Current address}

\begin{abstract}
We study the influence of transverse electric fields on the interfacial forces between a graphene layer and a carbon nanotube tip by means of atomistic simulations, in which a Gaussian regularized charge-dipole potential is combined with classical force fields. A significant effect of the field-induced electric charge on the normal force is observed. The normal pressure is found to be sensitive to the presence of a transverse electric field, while the friction force remains relatively invariant for the here-used field intensities. The contact can even be turned to have a negative coefficient of friction in a constant-distance scenario when the field strength reaches a critical value, which increases with decreasing tip-surface distance. These results shed light on how the frictional properties of nanomaterials can be controlled via applied electric fields.
\end{abstract}

\maketitle

\section{Introduction}
Scanning probe microscopy (SPM) is widely used in the characterization of nanomaterials \cite{Szlufarska2008a}. Concentration of electric charge in sharp edges is an important but unexpected factor in SPM experiments. It has been reported that electric charge can be locally induced in low-dimensional carbons by sensing concentrated charges in adjacent surfaces \cite{Poncharal-99,Wang2008a,Wang2009a}. If two neutral surfaces separated by a nanometric distance are charged, interfacial forces can be dramatically enhanced by orders of magnitude due to electrostatic interactions. Hence, the effects of applied electric fields or doping charges can be prominent in SPM measurements. Indeed, it has been shown that the adhesion and frictional forces are dictated by the electronic charge redistribution occurring due to the relative displacements of the two surfaces in contact \cite{Wolloch2018}. However, the mechanisms of electrostatic effects at nanostructured interfaces are so far poorly understood.

On the other hand, control of friction is critical for nanodevices with moving parts, due to the extreme surface-to-volume ratio of nanostructures \cite{Kendall1994}. Graphene is playing a central role in the design of many devices thanks to its superior mechanical strength, chemical inertness and extraordinary frictional properties distinct from bulk materials \cite{Verhoeven2004,Dienwiebel2004,Smolyanitsky2012,Egberts2014,Klemenz2014,Wang2019f,wang2018}. Precise control of the frictional proprieties of graphene is therefore of technological importance for many applications \cite{Berman2014}, especially for taming the mechanical motion of nanodevices \cite{Kim2007}. Recently, strain engineering methods have been proposed to control the friction of graphene on different types of substrates \cite{Wang2019,Yang2017}. Drummond demonstrate that a control of the global friction of a polyelectrolyte coating is possible by applying an alternating electric field \cite{Drummond2012}. Nonetheless, possible electrical means to the same end remain largely unexplored in contrast to the mechanical and chemical means.

With the motivations above, here we study the effect of transverse electric fields on the frictional properties of graphene probed by a carbon nanotube (CNT), using atomistic model that combines classic force fields with a Gaussian-regularized charge-dipole model \cite{Wang2007a,Yang2018}. A possible significant effect of the transverse electric field on the normal force is demonstrated below. 

\section{Methods}

\begin{figure}[htp]
\centerline{\includegraphics[width=8cm]{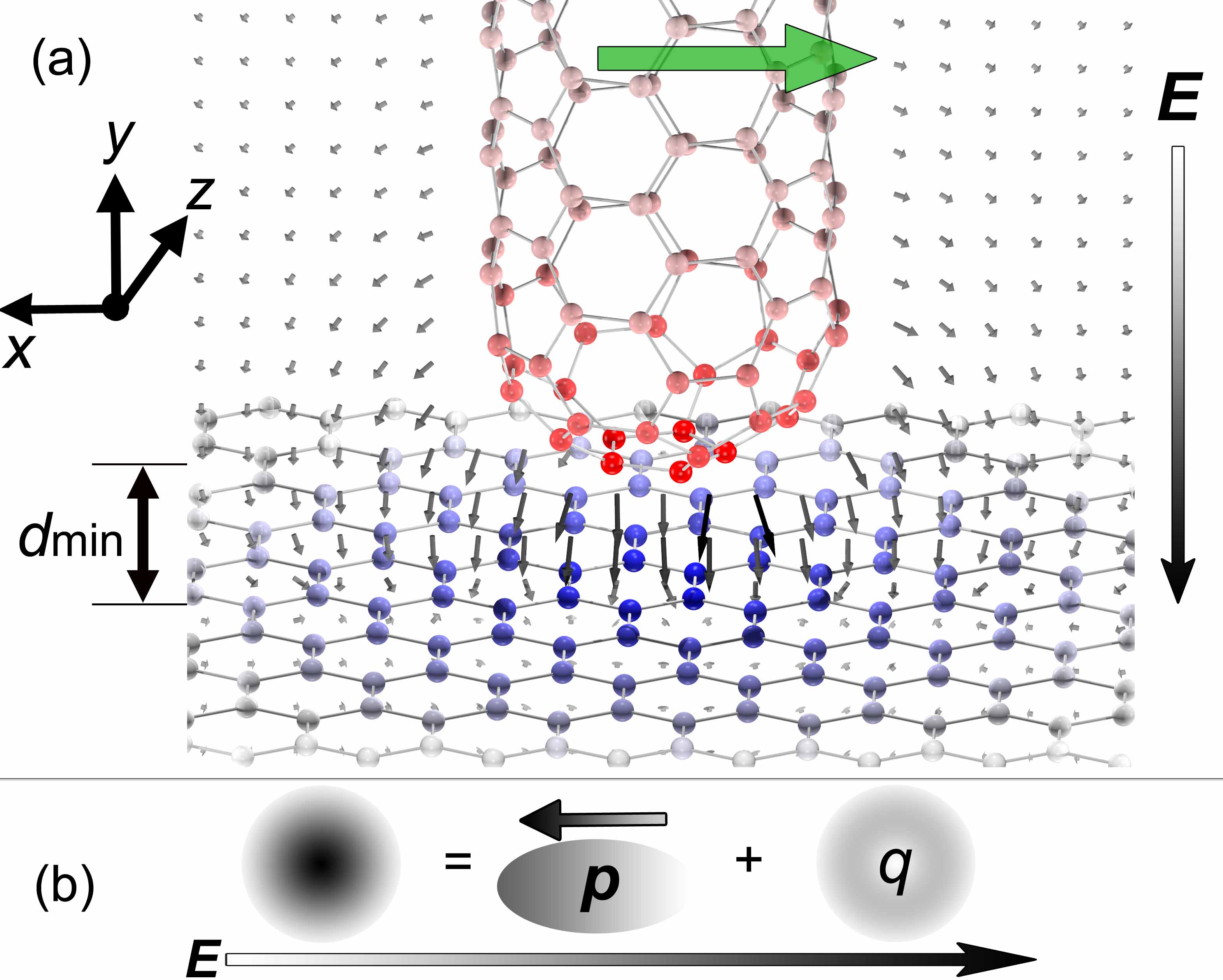}}
\caption{\label{F1}
(a) Simulation snapshot showing a CNT sliding along the zigzag direction over a graphene layer in a homogeneous electric field applied against the $y$ axis. The atoms are colored according to the density of induced electric charge (blue for negative and red for positive). The gray-scaled arrows are parallel to the local electric field, with their length and color representing the field strength. (b) Schematic for the \textit{qp} model in which each atom in an electric field is represented by a net electric charge $q$ plus an atomic dipole $\bm{p}$.}
\end{figure}

In our simulations, the tip of a capped (5,5) CNT ($68\;\mathrm{nm}$ long) is conducted to slide along the $-\bm{x}$ direction atop an infinite graphene layer in a constant-distance scheme, as shown in Fig.\,\ref{F1}\,(a). The atomistic configuration of the CNT tip is generated by integrating 5-7-7-5 defects in the hexagonal carbon lattice \cite{Reich2005,Zhu2016}, and is relaxed by the gradient descent method using the reactive empirical bond order (REBO) potential \cite{Brenner2002,Wang2019a,Wang2011a}. The graphene layer is infinite with a cell of about $6.6 \times 8.3$ nm in size with periodic boundary condition applied in both in-plane directions. Note that the cell of graphene should large enough to exclude the effect of periodic images. A homogeneous electric field is then applied vertically along the CNT axis normal to the graphene plane. Both the tip and the graphene layer are fixed rigid during the simulations, in order to reduce the requirement of the computational resource to an affordable level and to focus on the field effect without taking into account the influence of displacement rate, temperature and so forth. This approximation makes our simulations different from classical molecular dynamics \cite{Guo2015,Yang2015a,Qi2018,Wu2019}. The atomistic interaction potential $\varepsilon^{inter}$ between the tip and the graphene includes a van der Waals (vdW) term and an electrostatic (elec) term,

\begin{equation}
\label{eq1}
\varepsilon^{inter}=\varepsilon^{vdW}+\varepsilon^{elec}.
\end{equation}

$\varepsilon^{vdW}$ is provided by the GraFF force field, which extends the Lennard-Jones (LJ) potential that is known to underestimate the graphene surface energy corrugation \cite{Kolmogorov2000}. It is a sum of pair contributions,

\begin{equation}
\label{Eq2}
\varepsilon^{vdW}= \left \{ \begin{array}{ll}
0 &  \,\,\,\, 0<\theta<\frac{\pi}{4}\\
4 \epsilon \left[ \left( \frac{\sigma}{r_{ij}} \right)^{12} - \left( \frac{\sigma}{r_{ij}} \right)^{6} \right] \cos^{2}{2\theta} & \,\,\, \frac{\pi}{4}<\theta<\frac{3\pi}{4}\\
0 & \, \frac{3\pi}{4}<\theta<\pi 
\end{array} \right.
\end{equation}

\noindent where $r_{ij}$ is the distance between the atoms $i$ and $j$, $\epsilon$ stands for the depth of the potential well and $\sigma$ is the zero-potential interatomic distance. $\theta$ is the angle between $\bm{r}_{ij}$ and an in-plane covalent bond vector of the atom $i$. This angular dependence mimics a $\pi$-orbital in the region above the graphene where electron overlap occurs. The parameter values and benchmarks of the GraFF force field are provided in Ref.\citep{Sinclair2018}.

When a transverse electric field is applied along $-\bm{y}$, positive charges will shift to the tip to resist the application of an external field. In the case there is a graphene surface nearby, charge of opposite sign will be induced in the graphene near the tip, as shown in Fig.\,\ref{F1}\,(a). An adhesive electrical force is generated at the interface by Coulomb attraction between the accumulated and induced changes. This polarization effect is described by the Gaussian-regularized charge-dipole (\textit{qp}) model \cite{Wang2008,Wang2010a,Wang2009carbon,Wang2009d,Wang2010b} in which each atom is associated with an induced dipole and an amount of electric charge $q$ as shown Fig.\,\ref{F1}\,(b). The electrical potential can be expressed as

\begin{multline}
\label{Eq3}
\varepsilon^{elec} =\sum_{i=1}^N{q_i(\chi_i+V_i)}-\sum_{i=1}^N{\bm{p}_i\cdot\bm{E_i}}+
\sum_{i=1}^N{\sum_{\substack{j=i}}^N{q_i T^{i,j}_{q-q} q_j}}\\-\sum_{i=1}^N{\sum_{\substack{j=1}}^N{\bm{p}_i \cdot \bm{T}^{i,j}_{p-q} q_j}}-\sum_{i=1}^N{\sum_{\substack{j=i}}^N{\bm{p}_i \cdot \bm{T}^{i,j}_{p-p} \cdot \bm{p}_j}},
\end{multline}

\noindent where $\chi$ is the atomic electronegativity, $V$ and $\bm{E}$ stand for the external potential and the electric field, respectively. $T$ and $\bm{T}$ are the electrical interaction tensors (so-called vacuum propagators) between point charges and dipoles. The charge-charge, charge-dipole, dipole-dipole, charge-field, dipole-field interactions are taken into account in different terms. The charges and the dipoles are considered to be spherically symmetric, radially Gaussian, electronic charge distributions, in order to avoids the typical divergence problems called ``polarization catastrophes''. The \textit{qp} model remains the state-of-the-art method for large systems that are inaccessible to \textit{ab-initio} calculations. More details, including the parameterization and experimental validations are provided in our previous works \cite{Wang2007a,Wang2008a,Wang2007}.

\section{Results and Discussion}

\begin{figure}[htp]
\centerline{\includegraphics[width=8cm]{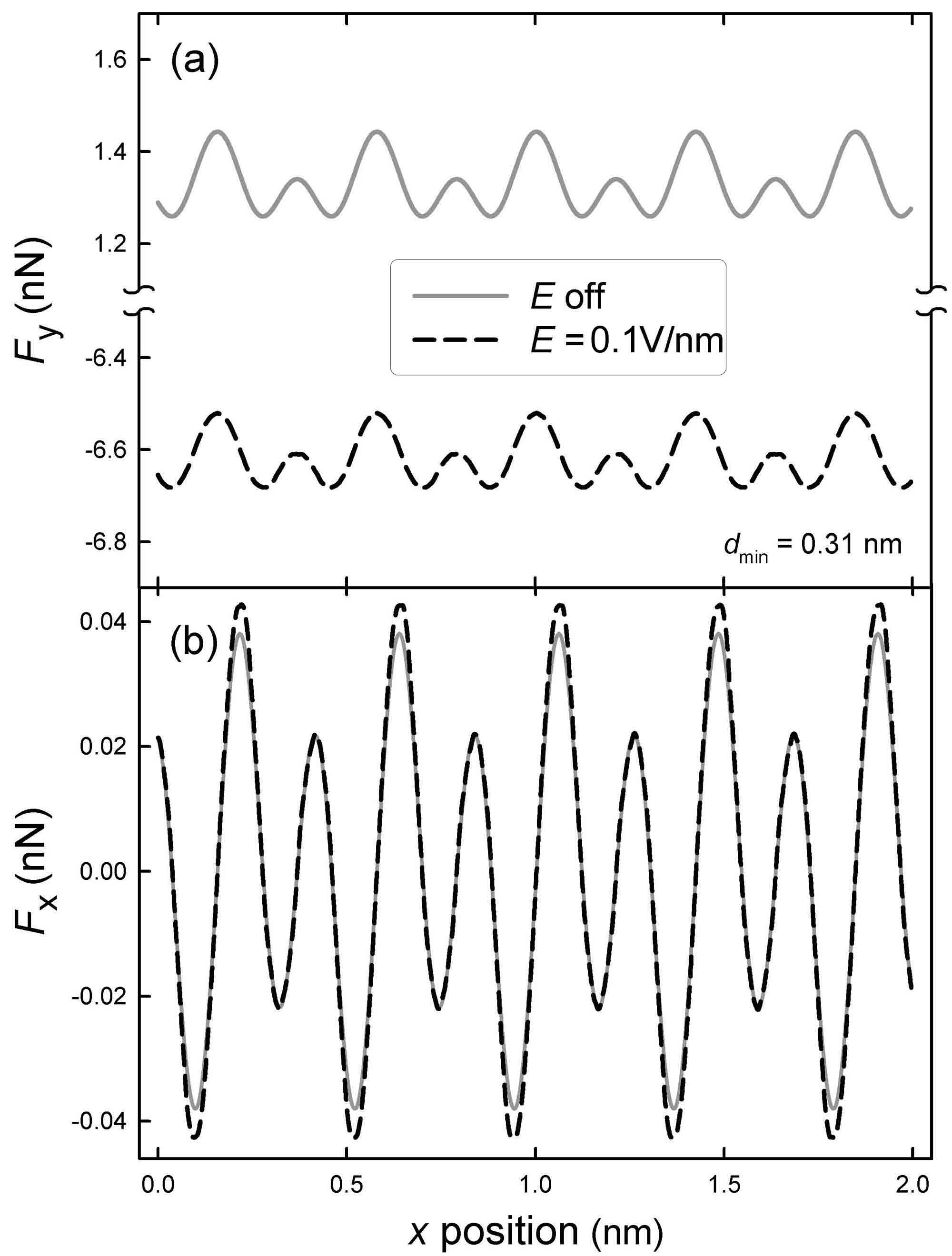}}
\caption{\label{F2}
Forces acting on the tip along the $y$ (a) or $x$ (b) directions \textit{versus} sliding distance for a minimum tip-graphene spacing of $d_{min}=0.31\;\mathrm{nm}$, under (dashed line) and without (solid line) a transverse electric field of $0.1\;\mathrm{V/nm}$.}
\end{figure}

The forces acting on the tip from graphene (equivalent to the negative of the external load at equilibrium) in the sliding and transverse (normal) directions are calculated as a function of the sliding distance of the tip in an electric field of $0.1\;\mathrm{V/nm}$, as shown in Fig.\,\ref{F2}. It is seen that the normal force $F_{y}$ oscillates for keeping a constant tip-graphene distance. When the electric field is applied, $F_{y}$ changes significantly and even points in the opposite direction, while the friction force remains almost invariant. i.e. the vdW force is cancelled out by the electrostatic force and the external load therefore becomes negative. This switches the external load (negative of the total sum of the vdW and the electrostatic forces) from positive (repulsive contact) to negative (adhesive contact). Note that an adhesive contact is defined as a contact that needs a negative external load to maintain a given tip-substrate distance. An adhesive contact would not necessarily lead to low friction force although adhesive contacts usually show higher friction forces than contacts with low adhesion.

Unlike the vdW force which can be either repulsive or attractive depending on the CNT-graphene distance, the electrostatic force can only be attractive regardless the direction of the applied transverse electric field, since the induced charges in the graphene always exhibit an opposite sign with respect to those on the tip. i.e. the effect of the transverse electric field can be considered as an extra adhesive force applied at the interface. This electrostatic attraction phenomenon has been observed in a transmission electron microscope experiment of Poncharal \textit{et al.} \cite{Poncharal-99}. It is known that the adhesive contact normal force can lead to a negative coefficient of friction \cite{Gao2004}, which has recently been reported by experiments \cite{Deng2012,Thormann2013} and simulations \cite{Mandelli2019} as a remarkable microscopic phenomenon. For a mult-layered graphitic system, the effect of a transverse electric field can be expected to be more pronounced due to further induced opposite-signed charges in the layers beneath the surface.

\begin{figure}[htp]
\centerline{\includegraphics[width=8cm]{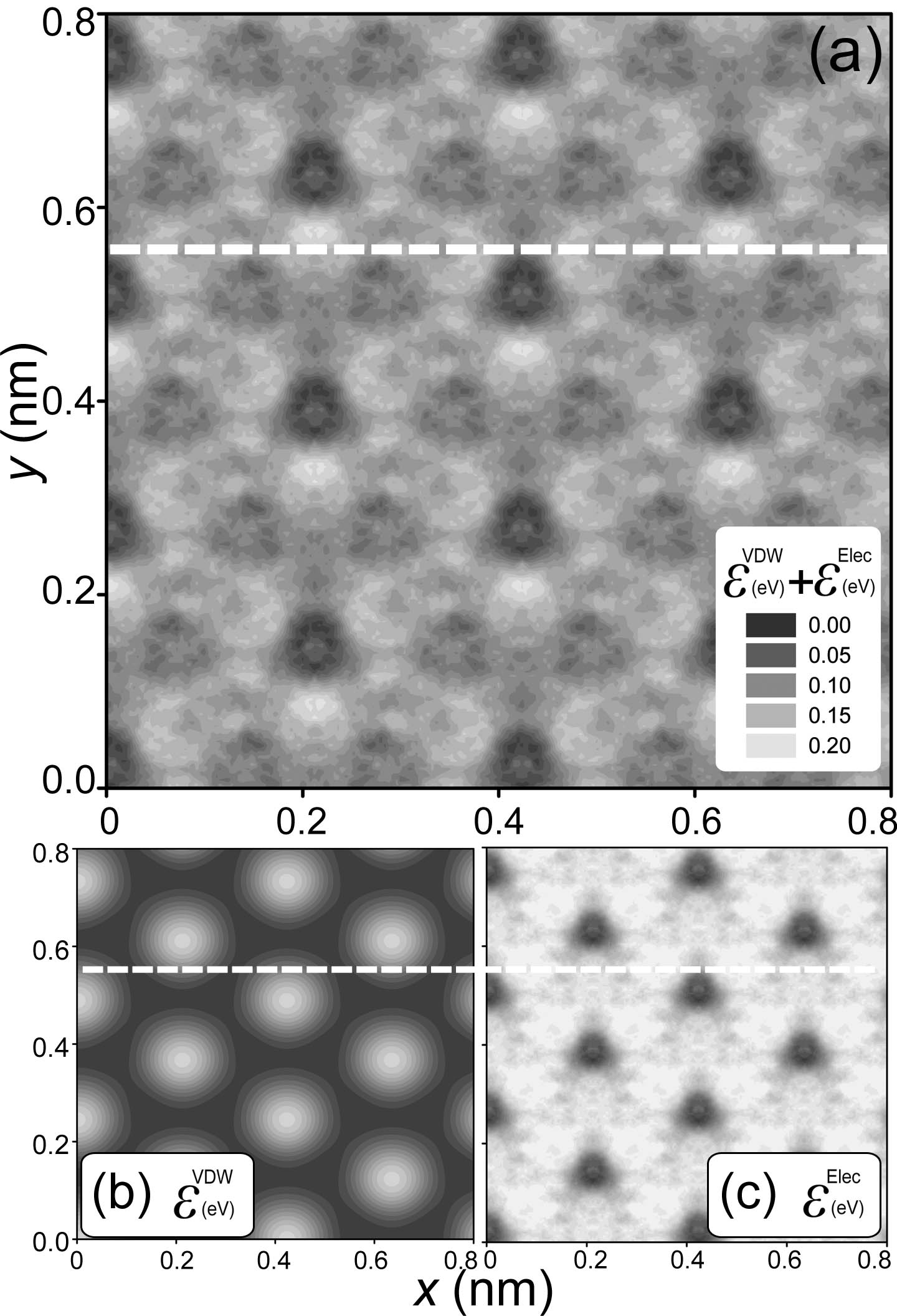}}
\caption{\label{F3}
Profiles of $\varepsilon^{inter}$ (a), $\varepsilon^{vdw}$ (b) and $\varepsilon^{elec}$ (c) for $d_{min}=0.31\;\mathrm{nm}$. The energy data are normalized to zero for the lowest values in each figure.}
\end{figure}

The $F_{y}$ and $F_{x}$ curves oscillate with a period of about $E=0.426\;\mathrm{nm}$ corresponding to the dimension of a periodic cell of graphene in the zigzag direction. This oscillation is also often measured in nanotribological experimentally \cite{Vilhena2016a,Dienwiebel2004}, and is correlated with the pattern of the potential energy surface (PES) shown in Fig.\,\ref{F3}. These distributions of potential energy are obtained by displacing the tip atop the graphene in discretized steps of $0.02\;$\AA \cite{Verhoeven2004}. e.g. when the CNT tip move from left to right as following the dashed line in Fig.\,\ref{F3}, oscillation of the forces will be induced by energy corrugation along this path. It can also been seen that, around the symmetrical points, $\varepsilon^{vdw}$ has relatively flat slopes. In contrast, the gradient of $\varepsilon^{elec}$ is much more marked. This explains why $F_{x}$ is only changed around the peaks in Fig.\,\ref{F2}\,(b) when the field is applied. Note that there is also a connection between the change of the PES with load and a negative coefficient of friction (CoF) for certain systems \cite{Righi2007}.

\begin{figure}[htp]
\centerline{\includegraphics[width=8cm]{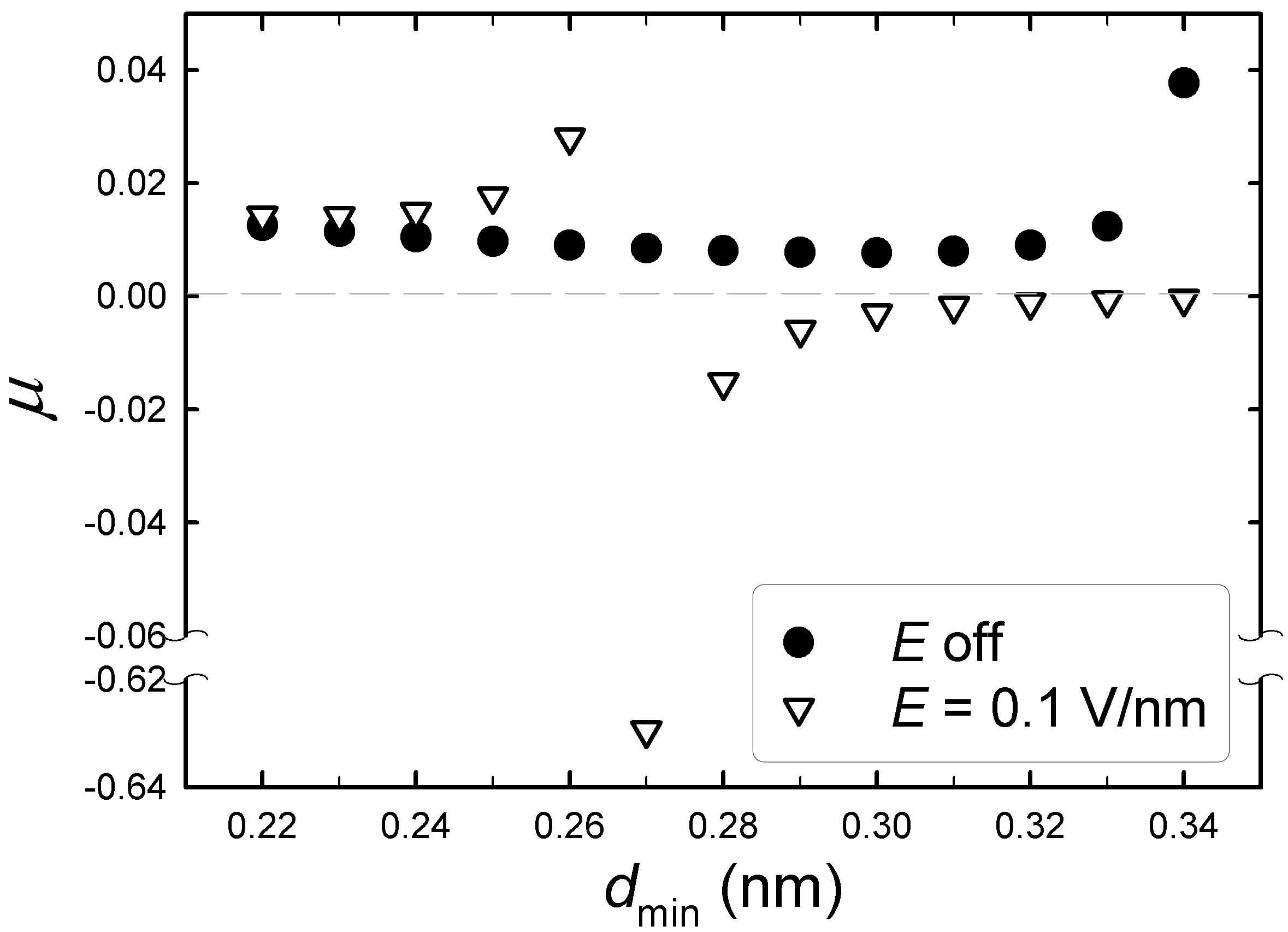}}
\caption{\label{F4}
CoF $\mu$ \textit{versus} $d_{min}$ in the absence (solid circles) and presence (empty triangles) of a transverse electric field $E=0.1\;\mathrm{V/nm}$.}
\end{figure}

The CoF $\mu$ is calculated by the method proposed by Zhong and Tom{\'a}nek \cite{Zhong1990} based on the measured forces $F_{y}$ and $F_{x}$, as shown in Fig.\,\ref{F4}. This method ignores all negative contributions to the friction force in an assumption of a particular type of stick-slip behavior similar to the case that the tip is pulled by a very strong spring in a typical setup of the Tomlinson model \cite{Tomlinson1929}. This method is often used for \textit{ab-initio} calculations where full relaxations and thermostating was prohibited by computational cost \cite{Sun2018}. It is observed that $\mu$ increases slightly with the increasing tip-graphene distance $d_{min}$ until reaching a critical distance, at which a maximum effect of the electric field on $\mu$ occurs where the contact becomes adhesive. The field effect shows a dependence on $d_{min}$ because the vdW and electrical forces exhibit different decay rates with increasing $d_{min}$. The magnitude of the electrical force becomes comparable to that of the vdW force around the critical point. This implies a striking fact: There must exist a value of $d_{min}$ or $E$ at which the attractive electrical force is exactly equal to the repulsive vdW force. At that point the normal load would reach zero and the CoF would diverge and become undefined.

\begin{figure}[htp]
\centerline{\includegraphics[width=8cm]{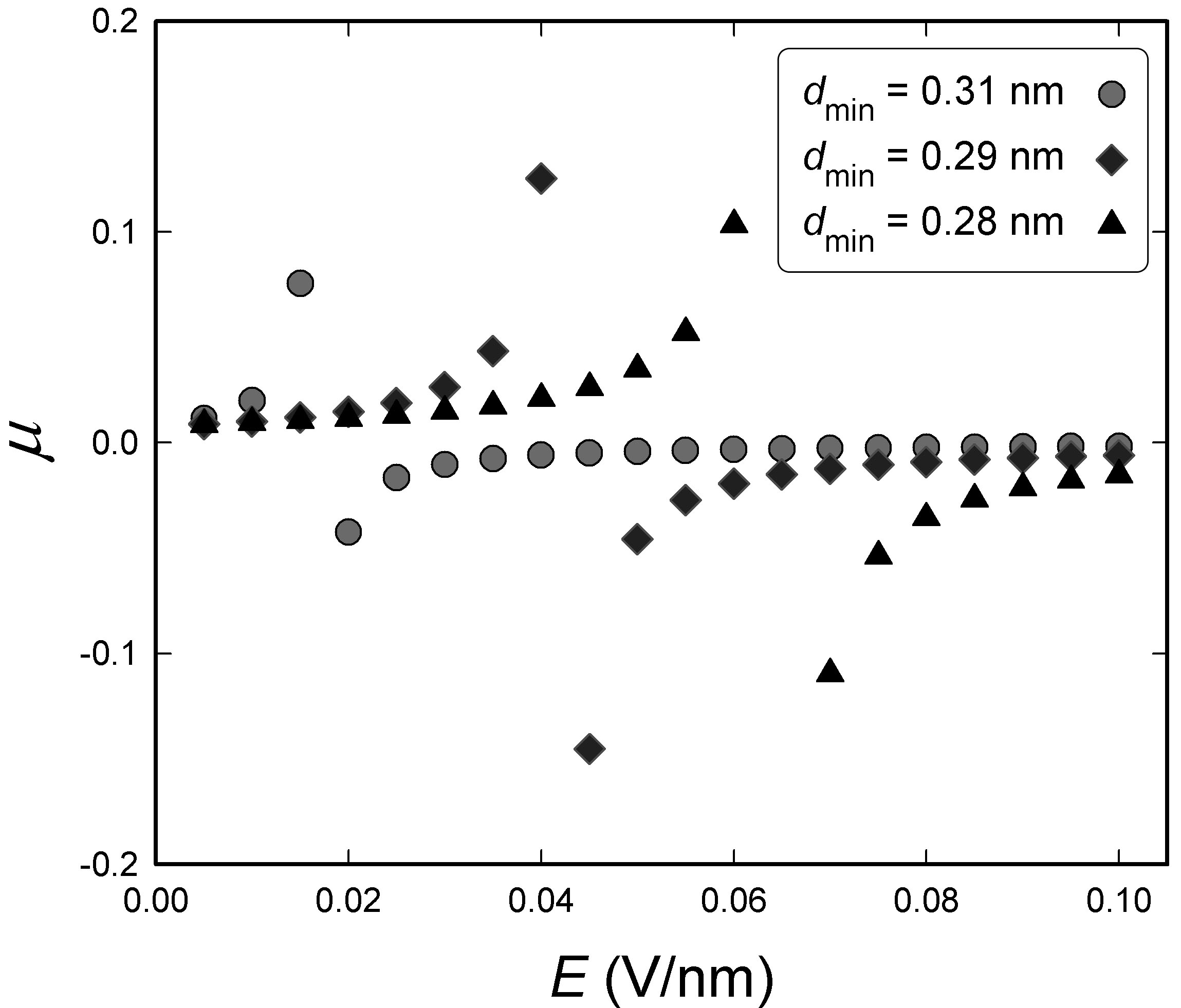}}
\caption{\label{F5}
CoF $\mu$ \textit{versus}field intensity $E$ for different tip-graphene spacings.}
\end{figure}

Fig.\,\ref{F5} shows how $\mu$ changes with the field intensity $E$. Under progressively more intense electric fields, $\mu$ first increases and then becomes negative when the field strength reaches a critical value $E_{c}$ . From marked positions of $E_{c}$ in Fig.\,\ref{F5}, an increase in $E_{c}$ can be observed with decreasing $d_{min}$, due to the fact that the electrical force grows less rapidly than the vdW repulsive force with decreasing $d_{min}$. When $E$ is increased beyond $E_{c}$, $\mu$ increases due to the enhanced adhesive force.

\section{Conclusions}
It is shown that transverse external electric fields exhibit dramatic effects on the frictional properties of graphene. The normal force at the interface is found to change significantly when fields are applied, while the friction force remains less sensitive for the field intensities explored here. It is observed that the CoF switches to negative when the field strength reaches a critical value, which is inversely correlated with tip-graphene distance. An issue of the divergence of the CoF is raised. Theoretically, the normal force can be even cancelled out at every instant if an alternating electric field is applied with carefully-chosen field strength in a frequency adapted to the profile of the potential energy according to the external load. Indeed, the knowledge of the PES is powerful for manipulating the friction properties of nanostructures.

Regarding the friction force, huge electric fields may be needed to produce significant effects. However, it is highly possible that such strong fields will induce important field-emission phenomena \cite{Purcell2002} and damage the CNT as well as the graphene before the friction can be measured. It should be mentioned that the system taken as an example here is rather artificial. For instance, the CNT is used here as the tip for its relatively small size, in experiments it is common to use much larger tips of different materials. Moreover, it is worth noting that the field strengths ($5-100\;\mathrm{V/{\mu}m}$) used here are lower than those used by previous simulations \cite{Guo2003,Wang2015} but still relatively high to be realized in common experiments. This is because our CNT is relatively short ($68\;\mathrm{nm}$); weaker fields will be needed in real SPM experiments often using micron-sized CNTs, since the induced electric force should be roughly proportional to the square of the length of the probe \cite{Wang2008}. 

\section*{Acknowledgments}
Qunyang Li at Tsinghua University, Harold Park at Boston University and Oded Hod at Tel-Aviv University are acknowledged for helpful discussions. This work is supported by the Guangxi Science Foundation (2018GXNSFAA138179) and the scientific research foundation of Guangxi University (XTZ160532).


\end{document}